\documentclass[sigconf]{acmart}

\acmConference[Conference acronym 'XX]{Make sure to enter the correct conference title from your rights confirmation email}{June 03--05, 2018}{Woodstock, NY}

\usepackage{amsmath,amsfonts,amssymb}
\usepackage{multirow}
\usepackage{makecell}
\usepackage[linesnumbered,ruled]{algorithm2e}
\usepackage{booktabs}
\usepackage[table,xcdraw]{xcolor}
\usepackage{color}
\usepackage{pifont}
\usepackage{adjustbox}

\begin{document}

\title{TriniMark: A Robust Generative Speech Watermarking Method for Trinity-Level Traceability}

\author{Yue Li}
\affiliation{%
  \institution{Huaqiao University}
 \country{}
}
\author{Weizhi Liu}
\affiliation{%
  \institution{East China Normal University}
 \country{}
}

\author{Kaiqing Lin}
\affiliation{%
  \institution{Shenzhen University}
 \country{}
}

\author{Dongdong Lin}
\affiliation{%
 \institution{Huaqiao University}
 \country{}
}

\author{Kassem Kallas}
\affiliation{%
  \institution{French National Institute of Health and Medical Research		
}
 \country{}
}





\begin{abstract}
Diffusion-based speech generation has achieved remarkable fidelity, increasing the risk of misuse and unauthorized redistribution. However, most existing generative speech watermarking methods are developed for GAN-based pipelines, and watermarking for diffusion-based speech generation remains comparatively underexplored. In addition, prior work often focuses on content-level provenance, while support for model-level and user-level attribution is less mature.
We propose \textbf{TriniMark}, a diffusion-based generative speech watermarking framework that targets trinity-level traceability, i.e., the ability to associate a generated speech sample with (i) the embedded watermark message (content-level provenance), (ii) the source generative model (model-level attribution), and (iii) the end user who requested generation (user-level traceability). TriniMark uses a lightweight encoder to embed watermark bits into time-domain speech features and reconstruct the waveform, and a temporal-aware gated convolutional decoder for reliable bit recovery. We further introduce a waveform-guided fine-tuning strategy to transfer watermarking capability into a diffusion model. Finally, we incorporate variable-watermark training so that a single trained model can embed different watermark messages at inference time, enabling scalable user-level traceability. Experiments on speech datasets indicate that TriniMark maintains speech quality while improving robustness to common single and compound signal-processing attacks, and it supports high-capacity watermarking for large-scale traceability.
\end{abstract}



\maketitle

\section{Introduction}
Generative speech models have rapidly advanced in recent years, enabling high-fidelity and natural-sounding synthetic speech that is increasingly indistinguishable from real recordings.
In particular, speech generation has been driven by two dominant paradigms —vocoder architectures based on Generative Adversarial Networks (GANs) ~\cite{goodfellow2014gan,kong2020hifigan} and diffusion models (DMs) ~\cite{ho2020ddpm, kong2020diffwave, lee2021priorgrad, chen2020wavegrad, chen2021wavegrad2}—which make large-scale, low-cost synthesis and customization practical.
While these advances benefit content creation and human–computer interaction, they also amplify security and societal risks in the speech domain, including voice impersonation and malicious forgery, unauthorized redistribution of synthetic speech, and disputes over the ownership and responsible use of deployed speech generators.
Consequently, there is an urgent need for proactive and verifiable protection mechanisms that enable reliable traceability for generated speech and the underlying generative models. Even in large-scale deployment, suspicious speech should be traceable to the user who generated it.

\begin{figure}[t]
    \centering
    \resizebox{0.98\linewidth}{!}{\includegraphics{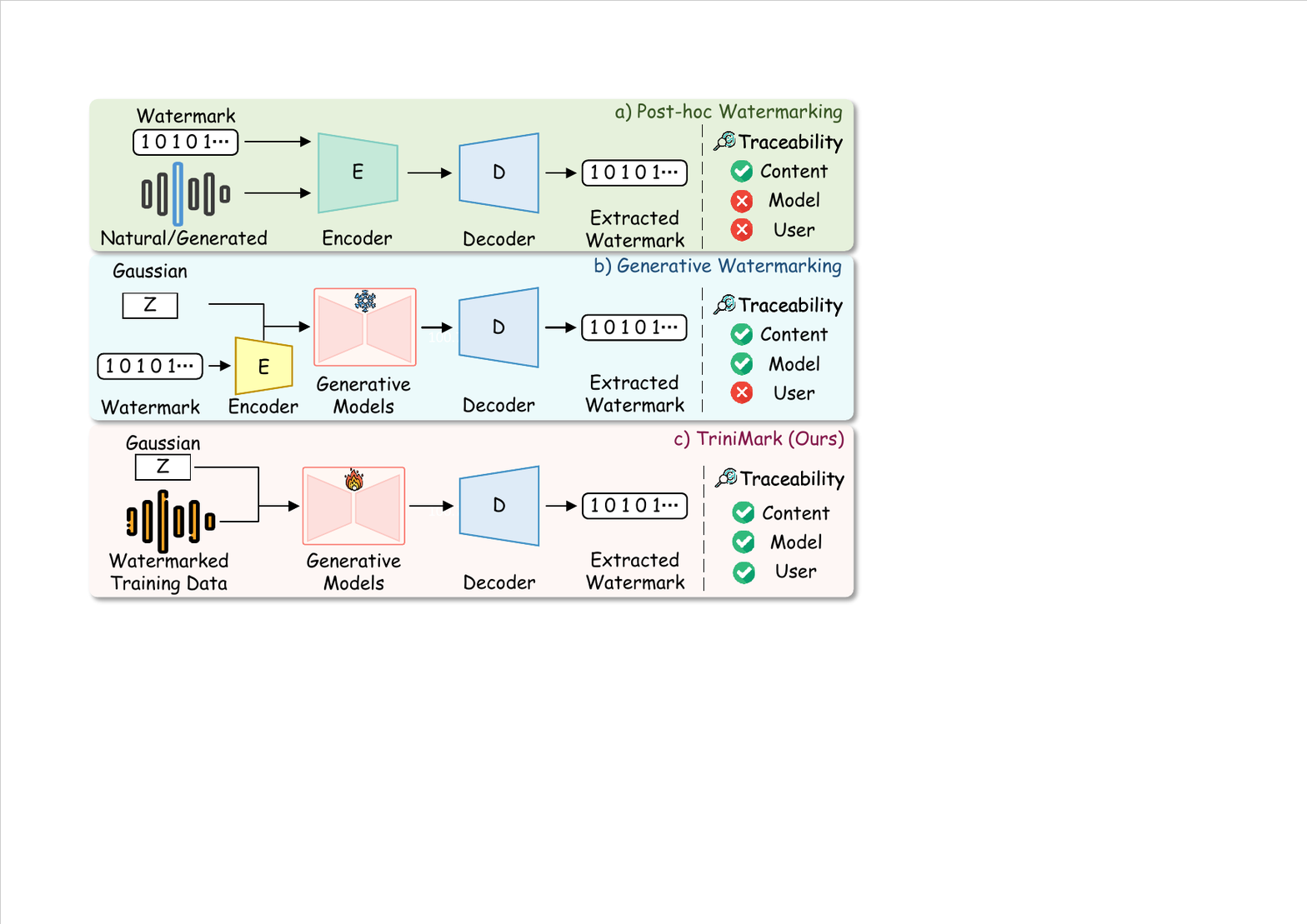}}
    \vspace{-0.3cm}
    \caption{The traceability scope of two speech watermarking paradigms and the proposed method. (a) Post-hoc watermarking embeds watermarks into speech after generation, focusing on content-level traceability. (b) Majority of generative watermarking guides the generator to synthesize watermarked speech, enabling traceability of both the content and the source model. (c) TriniMark further enables unified three-stage traceability across the generated speech, the source model, and the end user. 
    }
    \vspace{-0.4cm}
    \label{fig_overview}
\end{figure}

In response, watermarking is increasingly used as a proactive mechanism to provide verifiable provenance and traceability across modalities, with the image domain being the most mature. Broadly, prior image watermarking methods pursue traceability from two perspectives: (i) content watermarking, which embeds a verifiable identifier into the content to support authenticity checks and traceability after distribution ~\cite{zhu2018hidden,tancik2020stegastamp,fernandez2022ssl-wm,bui2023rosteals, wen2023treeing,yang2024gaussianshading}; and (ii) model watermarking, which embeds an identifiable signature into the generative model so that its outputs can serve as evidence of model ownership~\cite{yu2021artificial,fernandez2023stable,lin2024cycleganwm,asnani2024promark,li2021survey, kallas2023mixer,lin2025exploiting}. However, speech watermarking faces unique challenges due to its temporal structure and diverse channel distortions, making these image-oriented techniques non-trivial to directly transfer.


Against this backdrop, recent efforts have started to develop speech-specific watermarking methods tailored to generative speech systems. These deep-learning-based (DL-based) methods can be broadly categorized into the paradigms illustrated in Fig.~\ref{fig_overview}. \textit{Post-hoc speech watermarking} embeds a watermark into the generated speech after synthesis, treating the generator as a black box and operating directly on the output content~\cite{chen2023wavmark, roman2024audioseal, liu2024timebre, tong2024tfsw, li2024draw}. In contrast, \textit{generative speech watermarking} integrates watermarking into the speech generation process by injecting the watermark into the generative pipeline, such that the model synthesizes watermarked speech end-to-end during inference~\cite{cho2022attributable,juvela2024collaborative,zhou2024traceablespeech,san2024latentwm,cheng2024hifi,feng2025riwf,liu2024groot}.


Despite these advances, both paradigms provide only partial traceability. Post-hoc speech watermarking operates on the final waveform and is therefore inherently limited to content-level traceability: it can indicate whether a speech clip carries a watermark, but it is generally unable to reliably link the clip back to the source model or the responsible user under practical deployment. However, even generative speech watermarking has not yet achieved unified traceability across the content, model, and user levels. First, several representative works~\cite{cho2022attributable,juvela2024collaborative,san2024latentwm} follow zero-watermark designs and thus mainly support model-level identification rather than embedding an explicit message for different content and scalable multi-user tracing. Second, training-free schemes have been explored~\cite{liu2024groot}, yet the watermark is often fixed after training due to constraints imposed by the training protocol and the encoder–decoder design, making it difficult to distinguish speech generated by different users of the same deployed model. Third, recent attempts to extend traceability to users~\cite{cheng2024hifi,feng2025riwf,zhou2024traceablespeech} are limited by very small payloads (typically only tens of bits), which restricts scalability to a large user population. Moreover, these methods are frequently tailored to specific generators, such as ~\cite{cheng2024hifi} targets HifiGAN model~\cite{kong2020hifigan} and \cite{zhou2024traceablespeech} is designed for VALL-E architecture~\cite{wang2023valle}, leaving mainstream diffusion-based speech generation insufficiently covered.


To address these challenges, we propose \textbf{TriniMark}, a diffusion-based generative speech watermarking method built upon fine-grained feature transfer, enabling \textit{trinity-level traceability across the content, model, and user levels} (Fig.~\ref{fig_overview}).
TriniMark adopts a two-stage training pipeline. In Stage I, we design and pretrain a time-domain-aware watermark encoder–decoder that supports reliable, high-capacity message embedding and extraction on speech waveforms, forming a stable watermarking backbone for subsequent transfer.
In Stage II, we fine-tune the diffusion speech generator with a waveform-guided strategy that injects watermark capability into the generation process via fine-grained feature alignment. Specifically, the pretrained encoder is used to produce watermarked training samples, and the diffusion model is jointly optimized with the pretrained decoder to balance speech quality and extraction accuracy. Crucially, variable watermark messages are used throughout both stages, enabling the deployed generator to assign distinct watermark sequences to different users during inference and thus supporting scalable user-level traceability.


In a nutshell, our contribution can be summarized as:
\begin{itemize}
\vspace{-0.25cm}
    \item 
    \textbf{Trinity-level traceability for generative speech watermarking.} We propose TriniMark, a diffusion-based generative speech watermarking framework that enables unified traceability across the content, model, and user levels.

    \item 
    \textbf{Temporal-feature-guided two-stage training.} We design a two-stage training pipeline that transfers a time-domain watermark encoder–decoder into a diffusion generator via waveform-guided fine-tuning, achieving a favorable quality–robustness–capacity trade-off.

    \item 
    \textbf{Scalable user-level traceability via variable-watermark training.} By incorporating variable watermark messages in both training stages, TriniMark supports per-user watermark assignment at inference time, making user-level accountability feasible at scale.

    \item 
    \textbf{Extensive evaluation of quality–robustness–capacity trade-offs.} Experiments across multiple speech datasets and diffusion backbones show that TriniMark preserves perceptual speech quality while achieving high bit-recovery accuracy under both individual and compound distortions, and delivers high-capacity payloads (500 bps) that scale to large user populations.
    
\end{itemize}

\section{Related Work}
\subsection{Post-hoc Speech Watermarking}

Post-hoc speech watermarking refers to embedding and extracting an imperceptible message directly from speech content after it is generated or obtained, without modifying the upstream generative model. A common paradigm is to first transform speech into a stable representation (e.g., frequency spectra or waveform coefficients). and then employ a learnable embedding–extraction framework (typically an encoder-decoder architecture) to imprint a watermark that is recoverable under practical distortions.

Frequency domain methods are among the most widely studied. Leveraging the fine-grained feature of the short-time Fourier transform (STFT),
Pavlović et al.~\cite{pavlovic2022rsw-dnn} pioneered an encoder-decoder framework for speech watermarking. 
Building upon this line, O'Reilly et al.~\cite{or2024maskmark} incorporated Transformer~\cite{vaswani2017transformer}-based modeling with a multiplicative spectrogram mask to embed watermarks more effectively, 
while Chen et al.~\cite{chen2023wavmark} further improved the performance of watermarking via invertible neural networks.
To counteract voice cloning, Liu et al.~\cite{liu2024timebre} embedded repeated watermarks into STFT magnitude features. 
Beyond STFT-based designs, wavelet-domain watermarking has been explored for stronger robustness against re-recording.
Notably, Liu et al.~\cite{liu2023dear} employ the detail coefficients of Discrete Wavelet Transform (DWT) as the embedding space.
Although frequency domain watermarking is prevalent, there are also time-domain approaches that directly operate on speech waveforms.
Li et al.~\cite{li2025protecting} explored time-domain watermarking to protect singing voice content, while Roman et al.~\cite{roman2024audioseal}  focus on detecting AI-generated speech and localizing watermarks.

Overall, post-hoc methods primarily aim at content-level authentication of speech, which motivates complementary designs that integrate watermarking into the generation process for broader traceability.

\vspace{-0.5cm}
\subsection{Generative Speech Watermarking}

Generative speech watermarking integrates watermarking into the speech generation pipeline to synthesize watermarked speech end-to-end, enabling verifiable traceability for deployed speech generators. This paradigm is particularly relevant to modern speech generators that are deployed as shared services. Existing studies can be broadly grouped into (i) \textit{model-level identification}, where generated speech carries a detectable model-specific signature, and (ii) \textit{content-level message embedding}, where an explicit payload can be recovered from the generated speech under practical distortions.

Early efforts mainly target model-level identification rather than embedding an explicit payload for content-specific or multi-user tracing. Cho et al.~\cite{cho2022attributable} perform attribution by detecting model-specific signatures in generated speech. Juvela et al.~\cite{juvela2024collaborative} collaboratively train a neural vocoder with spoofing countermeasure models to improve detectability. 
San Roman et al.~\cite{san2024latentwm} watermark latent generative models via watermarking the training data to obtain decoder-robust detectability. These approaches are effective for model identification, yet they generally do not focus on embedding a content-dependent message or supporting scalable multi-user tracing, which limits their applicability to three-stage traceability.

More recent studies attempt to synthesize watermarked speech directly, but often face flexibility and scalability constraints. For instance, GROOT~\cite{liu2024groot} generates watermarked speech with a fixed diffusion backbone and auxiliary encoder/decoder, showing strong robustness against compound distortions. However, such training-free and parameter-fixed designs typically rely on a fixed watermark configuration after deployment, making it difficult to assign distinct watermark sequences to different users of the same model in practice. In contrast, HiFi-GANw~\cite{cheng2024hifi} fine-tunes a GAN vocoder with an extraction objective, and Feng et al.~\cite{feng2025riwf} incorporate watermark-related constraints into generator components. Nevertheless, existing attempts to extend traceability to the user side are frequently limited by small payloads (often only tens of bits), which fundamentally restricts scalability to large user populations.

Another practical limitation is that many generative speech watermarking methods remain architecture-specific, leaving the rapidly growing of diffusion-based speech generation insufficiently covered. For example, HiFi-GANw~\cite{cheng2024hifi} targets the HiFi-GAN vocoder, whereas TraceableSpeech~\cite{zhou2024traceablespeech} is designed for VALL-E~\cite{wang2023valle} style TTS architectures with frame-wise imprinting.
This leaves a practical open problem: enabling scalable user-level traceability with sufficiently high payloads while maintaining robustness and speech quality. Our work is motivated by this gap and investigates diffusion-based generative speech watermarking toward this goal.

\begin{figure*}
    \centering
    \resizebox{0.9\textwidth}{!}{\includegraphics{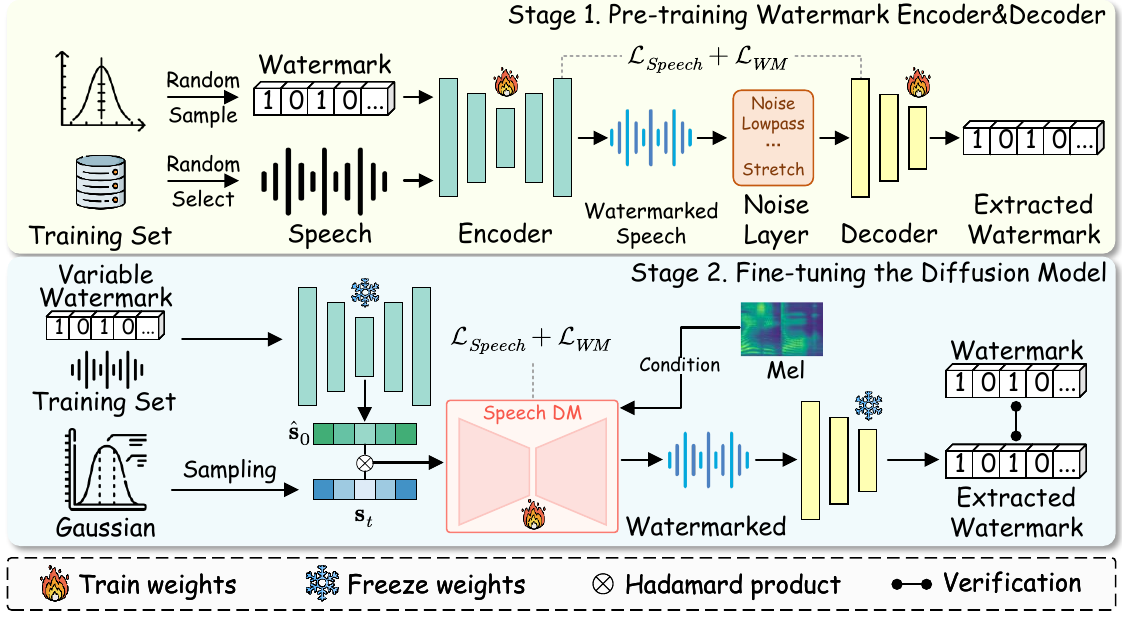}}
    \vspace{-0.3cm}
    \caption{Overview of TriniMark. TriniMark is trained in two stages: (1) pretraining a time-domain watermark encoder–decoder (trainable) with a robustness noise layer for reliable message embedding and extraction; and (2) waveform-guided fine-tuning of the diffusion vocoder (trainable), where the pretrained encoder and decoder are frozen to provide watermark priors and extraction supervision, enabling end-to-end synthesis of watermarked speech conditioned on mel-spectrograms. Variable watermark messages are used in both stages to enable per-user assignment at inference.}
    \label{fig_pipe}
    \vspace{-0.3cm}
\end{figure*}

\vspace{-0.50cm}
\section{Threat Model}





We consider a prompt-driven speech generation service in which a user prompt is first converted by an upstream front-end (such as an acoustic model or a large-model-based TTS component) into conditioning features, and a downstream vocoder synthesizes the time-domain waveform from these features. In this work, our diffusion-based watermarking method focuses on the vocoder stage, since it directly produces the distributable waveform and largely determines perceptual speech quality.

\underline{\textit{Capabilities and Goals of Users:}} The user is a legitimate client of the deployed system. The user can submit prompts and obtain synthesized speech waveforms, but does not possess privileged access to the model internals. The user is primarily concerned with speech quality and utility (e.g., naturalness, fidelity), and may redistribute the generated waveform, exposing it to downstream processing. In a shared-service setting, many users access the same vocoder, so user-level accountability requires assigning distinguishable watermark messages per user at inference.

\underline{\textit{Capabilities and Goals of Defenders:}} The defender represents the model owner or service provider who deploys the speech generator and aims to ensure verifiable traceability for generated speech in real-world distribution. The defender has full control over the watermarking design and deployment, including the embedding procedure integrated into the vocoder’s generation process, as well as the extraction and verification procedures used during audits. Importantly, while the overall system may include an upstream front-end, our threat model and protection objective are centered on the vocoder stage, since it produces the final waveform that is distributed and attacked.

In our deployment, watermark messages are also assigned by the defender rather than chosen by users. The defender maintains a server-side secret key $K$ and derives a per-user $l$-bit message $\mathbf{w}$ using a pseudorandom function, e.g., $\mathbf{w}=PRF_K(u||\tau)$, where $u$ is the user identifier and $\tau$ denotes a key index. The key $K$ is never revealed to users, and verification is performed on the defender side by extracting $\hat{\mathbf{w}}$ from a suspect clip and matching it against the expected messages (or their hashes) recorded by the service. Key rotation can be supported by updating $\tau$ while retaining prior versions for auditing within a bounded window.

\underline{\textit{Knowledge, Access and Goals of Adversaries:}} The adversary is an external party who obtains watermarked speech and attempts to undermine traceability. We focus on a black-box adversary: it has access only to the final watermarked waveform and may collect multiple examples, but has no access to the vocoder’s architecture, parameters, intermediate latents, watermark messages, or the defender’s verification system. The adversary cannot query gradients or manipulate model weights; its only capability is to apply post-processing to the waveform. The adversary’s goal is to weaken or desynchronize the watermark such that verification fails, while keeping the resulting speech perceptually acceptable. Under this black-box setting, we consider practical waveform-level manipulations that may arise in distribution channels or be applied intentionally: additive noise (e.g., Gaussian noise and pink noise), filtering (low-/high-pass), suppression, echo, and temporal modifications (e.g., time stretching and timing dither). We additionally consider compound attacks that sequentially combine multiple operations, capturing stronger but realistic watermark removal attempts.

An attack succeeds if it causes the defender’s verifier to reject a watermarked sample or reduces watermark recovery below a verification threshold, while maintaining acceptable perceptual quality. Finally, we clarify that we do not evaluate model-based re-synthesis attacks (e.g., passing the watermarked speech through another strong TTS model), which can substantially re-render the waveform and potentially eliminate embedded traces. Robust traceability under such re-generation settings is left for our future work.

\vspace{-0.3cm}
\section{Proposed Method}

TriniMark targets generative speech watermarking in a black-box setting by injecting an explicit, variable watermark message into diffusion-based waveform synthesis, rather than post-hoc editing the output speech or watermarking model parameters. Specifically, given a training speech 
$s$ and an $l$-bit message $\mathbf{w}$, our goal is to synthesize a watermarked waveform whose message can still be reliably recovered after post-processing. As depicted in Fig.~\ref{fig_pipe}, TriniMark adopts a two-stage transfer-learning pipeline: we first pretrain a lightweight time-domain watermark encoder–decoder to learn robust, high-capacity watermark priors, and then transfer these priors into the diffusion vocoder via waveform-guided fine-tuning. Crucially, variable-watermark training is used in both stages, enabling per-user message assignment at inference time and thus supporting scalable user-level traceability. Details of the two stages are provided in the following sections.

\subsection{Preliminaries}
TriniMark employs DDPM-based vocoders to synthesize watermarked speech. We briefly review the DDPM formulation for speech generation below.

In the diffusion process of DDPM, given natural speech $\mathbf{s}_0 \sim q_{data}(\mathbf{s}_0)$, the latent variable $\mathbf{s}_t$ is obtained by adding noise to it step by step, which follows the standard Gaussian distribution. This process follows a Markov chain:
\begin{equation}
\label{eq.forward}
    q(\mathbf{s}_t|\mathbf{s}_{t-1}) = \mathcal N(\mathbf{s}_t; \sqrt{1 - \beta_t} \mathbf{s}_{t-1}, \beta_t\mathbf{I}),
\end{equation}
\begin{equation}
    q(\mathbf{s}_{1:T}|\mathbf{s}_0) = \prod_{t=1}^T q(\mathbf{s}_t|\mathbf{s}_{t-1}),
\end{equation}
where $\beta_t \in (0,1)$ is the variance scheduled at time step $t$, and $\mathbf{I}$ is an identity matrix. Let $\alpha_t = 1 - \beta_t$, $\overline \alpha_t = \prod_{i=1}^t \alpha_i$, and $\mathbf{\epsilon} \sim \mathcal N(\mathbf{0}, \mathbf I)$. For any time step of $t$, by re-parameterization, the latent variable $\mathbf{s}_t$ can only be calculated from $\mathbf{s}_0$ and $\alpha_t$:
\begin{equation}
    \mathbf{s}_t = \sqrt{\overline{\alpha}_t} \mathbf{s}_0 + \sqrt{1-\overline{\alpha}_t} \mathbf{\epsilon}. 
\end{equation}
Therefore, the final diffusion process can be simplified to a single step. It can be represented as:
\begin{equation}
    q(\mathbf{s}_t|\mathbf{s}_0) = \mathcal N(\mathbf{s}_t; \sqrt{\overline{\alpha}_t} \mathbf{s}_0, (1-\overline \alpha_t)\mathbf I).
\end{equation}
 
The denoising process involves removing the noise from latent variable $\mathbf{s}_t$ by employing the prediction network $\mathbf{\epsilon}_\theta$ to estimate the noise added during the diffusion process step by step. This process $q(\mathbf{s}_{t-1} | \mathbf{s}_t, \mathbf{s}_0)$ also belongs to Gaussian distributed so that it can be computed as:
\begin{equation}
q(\mathbf{s}_{t-1} | \mathbf{s}_t, \mathbf{s}_0) = \frac{q(\mathbf{s}_t | \mathbf{s}_{t-1}) q(\mathbf{s}_{t-1}|\mathbf{s}_0)}{q(\mathbf{s}_t | \mathbf{s}_0)}
\end{equation}
According to Bayes' theorem. Unfolding this equation with Eq.~(\ref{eq.forward}) and combining like terms, we get
\begin{equation}
\label{reverse}
    q(\mathbf s_{t-1}\!|\mathbf s_t\!) =\!\mathcal N\!\big(\mathbf{s}_{t-1}\!; \frac{1\!}{\sqrt{\alpha_t}\!} (\mathbf s_t\! - \! \frac{1-\alpha_t\!}{\sqrt{1-\overline \alpha_t}} \mathbf{\epsilon}), (\frac{1 - \overline \alpha_{t-1}\!}{1 - \overline \alpha_t\!} \beta_t\!) \mathbf I ).
\end{equation}
Once the prediction network $\mathbf{\epsilon}_\theta$ has been well trained to predict noise $\epsilon$, the estimated speech can be obtained by $\mathbf{\epsilon}_\theta$:
\begin{equation}
    \mathbf s_{t-1} = \frac{1}{\sqrt{\alpha_t}} \bigg(\mathbf s_t - \frac{1-\alpha_t}{\sqrt{1-\overline \alpha_t}} \mathbf{\epsilon}_\theta(\mathbf s_t, t, c) \bigg) + \mathbf{\delta}_t \mathbf{z},
\end{equation}
where $\delta_t\mathbf{z}$ denotes the random noise, $\mathbf{z} \sim\mathcal N(\mathbf{0}, \mathbf{I})$, and $c$ is the mel-spectrogram.

The training of the prediction network aims to fit the noise $\epsilon$. Thus, the parameters $\theta$ need to be continuously learned and updated by maximizing the variational lower bound. Therefore, the final objective function can be simplified as:
\begin{equation}
\label{eq_train}
    \mathcal L_{simple} = \mathbb E_{t, \mathbf{s}_0, \epsilon} \big[ ||\epsilon - \epsilon_\theta(\sqrt{{\overline \alpha}_t} \mathbf{s}_0 + \sqrt{1 - {\overline \alpha}_t} \epsilon, t) ||^2 \big].
\end{equation}

\begin{figure}
    \centering
    \resizebox{0.98\linewidth}{!}{\includegraphics{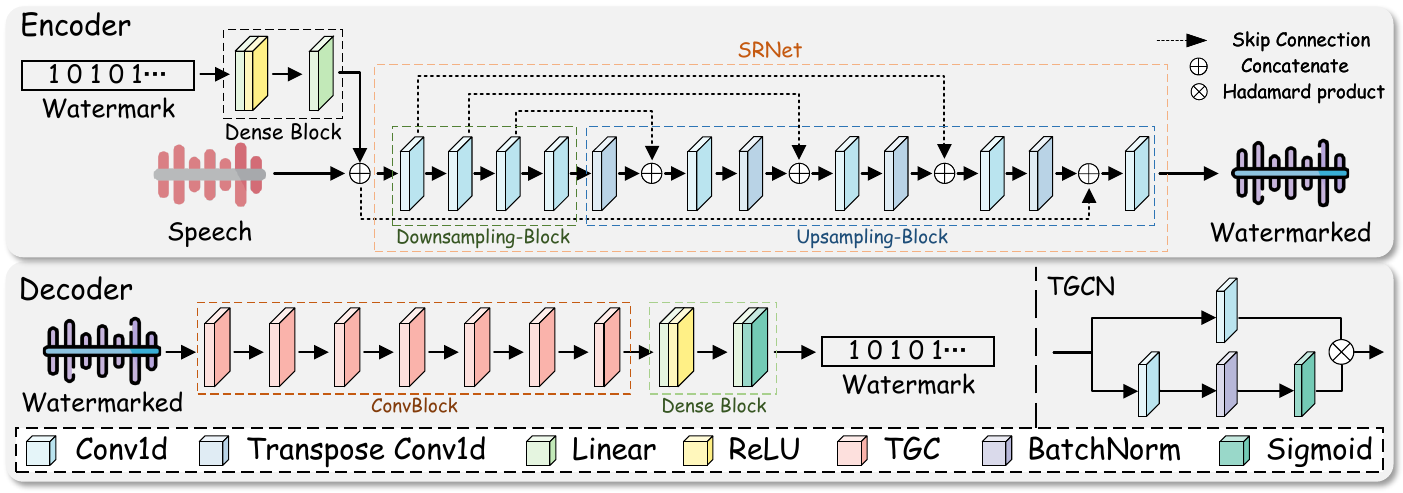}}
    \vspace{-0.3cm}
    \caption{The Detailed Architecture of Watermark Encoder and Decoder.}
    \vspace{-0.4cm}
    \label{fig_wm}
\end{figure}

\subsection{Pre-training Watermark Encoder and Decoder}
\subsubsection{Architecture of Watermark Encoder and Decoder}
We design a structure-lightweight watermark encoder $\mathcal{E}(\cdot)$ and a decoder $\mathcal{D}(\cdot)$ to learn transferable watermark priors for subsequent diffusion-model fine-tuning, as illustrated in Fig.~\ref{fig_wm}.
\textbf{The watermark encoder $\mathcal{E}(\cdot)$} consists of a DenseBlock followed by a Speech Reconstruction Network (SRNet). 
The DenseBlock maps the binary watermark into a compact embedding using two fully connected layers with ReLU, and SRNet injects this embedding into time-domain speech features and reconstructs the waveform. SRNet adopts a lightweight U-shaped design with a downsampling block and an upsampling block: the downsampling block contains four 1D convolutional layers, while the upsampling block comprises four 1D convolutional layers plus four 1D transposed convolutional layers. 
Different from frequency-domain reconstruction that typically relies on high-dimensional representations, our encoder operates on time-domain features and thus removes unnecessary activation and normalization layers to avoid losing low-dimensional information and to reduce computational overhead, improving both training efficiency and inference latency, following the lightweight design philosophy in MobileNetV2~\cite{sandler2018mobilenetv2}. 

\textbf{The watermark decoder $\mathcal{D}(\cdot)$} comprises a ConvBlock and a DenseBlock. The ConvBlock stacks seven deliberately designed Temporal-aware Gated Convolutional Networks (TGCNs), which adapt gated convolution to temporal patterns by redesigning Gated Convolutional Networks~\cite{dauphin2017gated_cnn}. Each TGCN has two branches: a linear 1D convolutional branch and a gated branch that applies 1D convolution followed by batch normalization and the Sigmoid function. The outputs of the two branches are fused via a Hadamard product to selectively pass watermark-related temporal cues. The resulting features are then aggregated by the DenseBlock for bit-wise watermark recovery.
For robustness, we introduce a noise layer $\mathbf{N}(\cdot)$ that simulates common post-processing distortions during training.

\subsubsection{Pipeline of Pre-training}
\label{sec_jointo}
This stage pretrains the watermark encoder–decoder to learn transferable watermark priors with both high-fidelity reconstruction and robust bit recovery.
Given a cover speech $\mathbf{s} \in \mathbb{R}^{\mathbf{n}}$ ($\mathbf{n}=C \times L$) and an $l$-bit watermark $\mathbf{w} \in \{0, 1\}^{l}$ (where $C$ represents channels and $L$ denotes the length of the speech), the encoder $\mathcal{E}(\cdot)$ first maps $\mathbf{w}$ into a compact embedding through the DenseBlock, and then injects it into time-domain speech features via SRNet to reconstruct the watermarked waveform $\hat{\mathbf{s}}$ ($\hat{\mathbf{s}}=\mathcal{E}(\mathbf{s},\mathbf{w})$). 
To improve robustness under real-world distribution,
we insert a noise layer $\mathbf{N}(\cdot)$ between the encoder and decoder, so the decoder $\mathcal{D}(\cdot)$ learns to recover the watermark from the attacked waveform $\mathbf{N}(\hat{\mathbf{s}})$ rather than from  $\hat{\mathbf{s}}$ itself. 
\begin{equation}
    \hat{\mathbf{w}} = \mathcal{D}(\mathbf{N}(\mathcal{E}(\mathbf{s}, \mathbf{w}))).
\end{equation}

A key design in our pre-training is \textit{variable-watermark training}, which prevents the encoder–decoder from overfitting to a fixed message and enables generalization to arbitrary watermarks. Concretely, we assign one pseudorandom $l$-bit watermark sequence per mini-batch and use that same sequence for all samples within the batch. Across mini-batches in the same epoch, different batches receive different watermark sequences. Moreover, at the beginning of each epoch, we regenerate a fresh pool of watermark sequences and reassign them to mini-batches, so the training procedure continuously exposes the model to new messages throughout optimization.

The pre-training objective jointly optimizes watermark recovery accuracy and speech fidelity.
For watermark extraction, we employ binary cross-entropy (BCE) as a constraint:
\begin{equation}
\label{eq_wm}
    \mathcal{L}_{W} = - \sum_{i=1}^{l} w_i \log {\hat{w}_i} + (1-w_i) \log(1-\hat{w}_i).
\end{equation}
Regarding the quality of watermarked speech, we first utilize the mel-spectrogram loss $\mathcal{L}_{Mel}$ to constrain the distance between the cover speech $\mathbf{s}$ and the watermarked speech $\hat{\mathbf{s}}$.
\begin{equation}
    \mathcal{L}_{Mel} = \mathbb{E}_{\mathbf{s}, \hat{\mathbf{s}}} \big[ || \phi(\mathbf{s}) - \phi(\hat{\mathbf{s}}) ||_1 \big],
\end{equation}
where $\phi(\cdot)$ represents the transformation of mel-spectrograms and $||\cdot||$ denotes the $L_1$ norm. Furthermore, we employ the logarithmic STFT magnitude loss $\mathcal{L}_{Mag}$ for optimization. 
\begin{equation}
    \mathcal{L}_{Mag} = || \log(\mathbf{STFT}(\mathbf{s})) - \log(\mathbf{STFT}(\hat{\mathbf{s}})) ||_1,
\end{equation}
where $\mathbf{STFT}(\cdot)$ denotes the transformation of STFT magnitude.
The entire loss of pre-training can be defined as:
\begin{equation}
    \mathcal{L}_{Pre} = \gamma_{mel} \mathcal{L}_{Mel} + \gamma_{mag} \mathcal{L}_{Mag} + \gamma_{w} \mathcal{L}_{W},
\end{equation}
where $\gamma_{mel}$, $\gamma_{mag}$, $\gamma_{w}$ are trade-off hyper-parameters for the speech fidelity terms and the watermark extraction term.

\subsection{Fine-tuning the Diffusion Models}
\subsubsection{Fine-tuning Strategy}
The standard training objective of diffusion vocoders constrains the predicted noise $\epsilon_\theta(\cdot)$ to match the injected noise $\epsilon$ (see Eq.~(\ref{eq_train})). 
Directly fine-tuning a diffusion model under this noise-matching objective, however, is not well aligned with watermark transfer: the watermark-related constraint tends to converge slowly, and overly emphasizing noise prediction may also degrade the vocoder’s generative capability, resulting in reduced speech quality.

To address this issue, we propose a novel waveform-guided fine-tuning strategy (WGFT). Instead of constraining $\epsilon_\theta(\cdot)$ and $\epsilon$, WGFT constrains the training speech $\mathbf{s}_0$ and the watermarked speech $\hat{\mathbf{s}}_0^{wm}$ obtained through the complete sampling process. This waveform-level guidance provides a more direct supervision signal for watermark transfer than noise-level matching, which empirically stabilizes optimization.
Concretely, we guide fine-tuning by minimizing the discrepancy between the ground-truth speech waveform $\mathbf{s}_0$ and its corresponding synthesized watermarked waveform $\hat{\mathbf{s}}_0^{wm}$. The rationale is that the pretrained diffusion vocoder already captures strong priors for noise prediction, and fine-tuning should focus on transferring watermark-ability while preserving synthesis fidelity.  

Following Section~\ref{sec_jointo}, we define a speech-quality constraint using the mel-spectrogram loss $\mathcal{L}_{Mel}$ and the log STFT magnitude loss $\mathcal{L}_{Mag}$:
\begin{equation}
    \mathcal{L}_{Speech} = \psi_{mel} \mathcal{L}_{Mel} + \psi_{mag} \mathcal{L}_{Mag},
\end{equation}
where $\psi_{mel}$ and $\psi_{mag}$ balance the two terms. For watermark recovery, we continue to employ the BCE loss in Eq.~(\ref{eq_wm}). 
The final loss for fine-tuning is defined as:
\begin{equation}
    \mathcal{L}_{FT} = \lambda_{speech} \mathcal{L}_{Speech} + \lambda_{w} \mathcal{L}_W,
\end{equation}
where $\lambda_{speech}$ and $\lambda_{w}$ are hyper-parameters to balance the two terms. In this way, WGFT encourages the diffusion vocoder to generate watermarked speech end-to-end while maintaining a favorable balance between speech quality and watermark extractability.

\newcommand{\lIfElse}[3]{\lIf{#1}{#2 \textbf{else}~#3}}
\begin{algorithm}[t]
\caption{Waveform-Guided Fine-tuning Pipeline.}
\label{algo}
\SetKwInOut{Input}{Input}
\SetKwInOut{Output}{Output}
\SetKwComment{Clean}{}{}

\Input{Training speech $\mathcal{D}_{train}$, mel-spectrogram $c$, pretrained watermark encoder $\mathcal{E}(\cdot)$ and decoder $\mathcal{D}(\cdot)$, variable watermark $\mathbf{w}$, noise schedule $\{\alpha_t\}_{t=1}^T \space \text{and}\space \{\bar{\alpha}_t\}_{t=1}^T$, and diffusion steps $T$.}

\Repeat{converged}{
    $\mathbf{s}_0 \sim \mathcal{D}_{train}$\;
    
    $\hat{\mathbf{s}}_{T}^{wm} \sim \mathcal{N}(0, \mathbf{I})$\;
    
    \For{$t \leftarrow T, \dots, 1$}{
        \lIfElse{$t > 1$}{$\mathbf{z} \sim \mathcal{N}(0, \mathbf{I})$}{$\mathbf{z} \leftarrow 0$}

         $\hat{\mathbf{s}}_{t-1}^{wm} \leftarrow \frac{1}{\sqrt{\alpha_t}} \big(\hat{\mathbf{s}}_t^{wm} - \frac{1-\alpha_t}{\sqrt{1-\overline \alpha_t}} \mathbf{\epsilon}_\theta(\hat{\mathbf{s}}_t^{wm}, t, c) \big) + \delta_t \mathbf z$\;
        
        \If{$t = 3$}{
        \tcp {Watermark Embedding}

            $\hat{\mathbf{s}}_0 \leftarrow \mathcal{E}(\mathbf{s}_0, \mathbf{w})$ \Clean*[r]{$\mathbf{w} \in \{0, 1\}^l$} 
        
            $\hat{\mathbf{s}}_t^{wm} \leftarrow \hat{\mathbf{s}}_t^{wm} \odot \hat{\mathbf{s}}_0$ \Clean*[r]{\tcp{$\odot$ Hadamard Product}}
        }
        
    }

    \Return $\hat{\mathbf{s}}_0^{wm}$\;

    \tcp{Watermark Extraction}
    $\hat{\mathbf{w}} \leftarrow \mathcal{D}(\hat{\mathbf{s}}_0^{wm})$\; 
    
    Take gradient descent step on: \\
    $\nabla_\theta \bigg( 
    || \phi(\mathbf{s}_0) - \phi(\hat{\mathbf{s}}_0^{wm}) ||_1 + || \log(\mathbf{STFT}(\mathbf{s}_0)) - \log(\mathbf{STFT}(\hat{\mathbf{s}}_0^{wm})) ||_1 - \sum_{i=1}^{l} \big(w_i \log {\hat{w}_i} + (1-w_i) \log(1-\hat{w}_i) \big)
    \bigg)$\;
}
\end{algorithm}

\subsubsection{Pipeline of Waveform-Guided Fine-tuning}
We fine-tune a DDPM-based vocoder with WGFT under the standard conditional diffusion setting, where the model is conditioned on the mel-spectrogram $c$ and trained using speech waveform $\mathbf{s}_0$.
Given the training speech $\mathbf{s}_0$ and the watermark $\mathbf{w}$, the pretrained encoder $\mathcal{E}(\cdot)$ embeds $\mathbf{w}$ into $\mathbf{s}_0$ to produce a watermarked target $\hat{\mathbf{s}}_0$, which is then diffused to obtain the Gaussian latent $\mathbf{s}_t$ and is used to train the diffusion vocoder.
The vocoder output $\hat{\mathbf{s}}_{0}^{wm}$ is fed into the pretrained decoder $\mathcal{D}(\cdot)$ to recover the watermark $\hat{\mathbf{w}}$, thereby providing waveform-level supervision for watermark transfer. In addition, we follow the variable-watermark training in Section~\ref{sec_jointo} by assigning one pseudorandom message per mini-batch and refreshing the message pool every epoch. The entire process of fine-tuning is defined as:
\begin{equation}
    \hat{\mathbf{s}}_{0}^{wm} = \mathcal{G}_D(\mathcal{E}(\mathbf{s}_0, \mathbf{w})),
\end{equation}
\begin{equation}
    \hat{\mathbf{w}} = \mathcal{D}(\hat{\mathbf{s}}_{0}^{wm}) \in \mathbb{R}^l,
\end{equation}
where $\mathcal{G}_D$ denotes the diffusion vocoder.
The overall procedure of fine-tuning is detailed in Algorithm~\ref{algo}.

\subsection{Watermark Verification}
The detection of the watermark $\mathbf{w}$ can be treated as a rigorous hypothesis test problem \cite{lin2024cycleganwm}. On this side, a verifier makes the decision $d$ regarding whether the watermark exists or not according to a given threshold. In this process, the verifier may make two types of errors under two hypotheses: the null hypothesis $\mathcal{H}_0$ represents the sample/model does not contain the watermark $\mathbf{w}$, and the alternative hypothesis $\mathcal{H}_1$ represents the sample/model contains $\mathbf{w}$.


The error of FPR can be constrained with a given threshold. The recovered watermark $\hat{\mathbf{w}}$ and the predefined watermark $\mathbf{w}$ are compared bitwise, and the number of matching bits $k$ follows the binomial distribution
\begin{equation}
\label{eq:threshold}
    \mathbb{P}(K=k) = \binom{l}{k}p^k(1-p)^{l-k},
\end{equation}
where $p$ is the probability. (for example, $p=0.5$ under the hypothesis $\mathcal{H}_0$). With a given FPR and the length of the watermark $l$, one has a threshold $th=k$ by solving the equation $\text{FPR}=\mathbb{P}(K)$.
After obtaining the threshold $th$, the verifier can make the decision whether that sample/model contains the watermark or not, based on bitwise accuracy $k/l$, e.g., the sample/model is watermarked if $k\geq th$, or otherwise. Then, given an FPR, choose the smallest $th$ such that $\mathbb{P}(K\ge th\mid \mathcal{H}_0)\le \mathrm{FPR}$.


In the hypothesis test, one should also consider the confidence of the decision made by the verifier. As given in \cite{lin2024cycleganwm}, if the verifier targets a $\text{FPR}=0.5\%$ with confidence 95\%, 768 test cases are needed for verification. 


\section{Experimental Analysis}
In this section, we validate the proposed TriniMark through comprehensive experiments on fidelity, capacity, robustness, and compare it with state-of-the-art (SOTA) methods. All results are reported across multiple datasets and diffusion vocoders.

\subsection{Experimental Setup}
\subsubsection{Datasets and Baseline}
\label{sec_dataset}
We conducted experiments using the LJSpeech~\cite{Ito2017ljspeech}, LibriTTS~\cite{zen2019libritts}, and LibriSpeech~\cite{panayotov2015librispeech} datasets. Concretely, LJSpeech is a single-speaker dataset containing nearly 24 hours of speech. LibriTTS and LibriSpeech are multi-speaker datasets containing approximately 585 hours and 1000 hours of data, respectively. We downsampled the speech in LibriTTS and LibriSpeech to 22.05 kHz. Since LJSpeech is already sampled at 22.05 kHz, no resampling was performed. To better assess the performance of the proposed method, we segmented all speech to a length of one second.
Furthermore, we utilize WavMark~\cite{chen2023wavmark}, AudioSeal~\cite{roman2024audioseal}, TimbreWM~\cite{liu2024timebre}, and Hifi-GANw~\cite{cheng2024hifi} as baselines for comparison.

\subsubsection{Evaluation Metrics}
\label{sec_metrics}
We evaluated the performance of our method with different objective evaluation metrics. 
Short-Time Objective Intelligibility (STOI) \cite{taal2010stoi} predicts the intelligibility of speech. 
The mean opinion score of listening quality objective assesses speech quality based on the Perceptual Evaluation of Speech Quality (PESQ)~\cite{recommendation2001mosl}. 
We also conducted evaluation metrics using Structural Similarity Index Measure (SSIM) \cite{wang2004ssim} and Mel Cepstral Distortion (MCD) \cite{kubichek1993mcd}. 
SSIM is a metric typically used for image quality assessment, which has also been adapted to the mel-spectrogram of speech for evaluation. 
MCD measures the reconstruction distortion of speech signals in terms of mel-frequency cepstral coefficients. 
Bit-wise accuracy (ACC) is employed to evaluate the accuracy of watermark extraction.


\vspace{-0.1cm}
\subsection{Implementation Details}
\subsubsection{Model Settings}
1) \textit{Watermark encoder and decoder.} In the Encoder, the first Fully Connected (FC) layer of the DenseBlock receives the watermark length as input and produces an output dimension of 512. The second FC layer generates an output dimension corresponding to the length of the speech signal. The SRNet employs 1D convolutional layers in its downsampling block, with a kernel size of 3, stride of 2, and padding of 2 for all layers except the final layer, which utilizes a padding of 1. In the upsampling block, all 1D convolutional layers maintain a kernel size of 3, stride of 1, and padding of 1. Furthermore, all 1D transposed convolutional layers have a kernel size of 3, stride of 2, padding of 2, and output padding of 1, except the first layer, which employs a padding of 1. 
Regarding the Decoder, each 1D convolutional layer in the TGCN is characterized by a kernel size of 3, stride of 2, and padding of 1. The first FC layer of the DenseBlock receives the feature length extracted by the ConvBlock as the input dimension, with an output dimension of 512. Meanwhile, the second FC layer produces an output dimension corresponding to the watermark length.
2)~\textit{Diffusion model}. We evaluate TriniMark on three representative diffusion vocoders: DiffWave~\cite{kong2020diffwave}, PriorGrad~\cite{lee2021priorgrad}, and WaveGrad~\cite{chen2020wavegrad}. DiffWave and WaveGrad are widely used accelerated-sampling vocoders; in particular, WaveGrad employs a gradient-based sampler inspired by Langevin dynamics for waveform synthesis. PriorGrad further improves efficiency by adopting an adaptive, data-driven prior conditioned on the input features.

\subsubsection{Training Settings}
1) \textit{Pre-training watermark encoder and decoder.} We jointly train the encoder and decoder using Adam~\cite{kingma2014adam} with a learning rate of 2e-4 for 80 epochs and a batch size of 16. 
We adopt a staged weighting strategy that first emphasizes watermark extraction by setting $\gamma_w=1$, $\gamma_{mel}=0$, and $\gamma_{mag}=0$. After $\mathcal{L}_W$ reaches a preset threshold, we switch on the speech-fidelity constraints by updating $\gamma_{mel}=0.9$ and $\gamma_{mag}=0.1$. 
2) \textit{Fine-tuning the diffusion models.} We fine-tune the diffusion vocoders using AdamW~\cite{loshchilov2018adamw} with a learning rate of 2e-4 for 20 epochs and a batch size of 2. Similarly, we first prioritize watermark extraction with $\lambda_w=1$ and $\lambda_{speech}=0$. And we set $\lambda_{speech}=1$ once $\mathcal{L}_W$ falls below the threshold. For speech loss $\mathcal{L}_{Speech}$, we use $\psi_{mel}=0.2$ and $\psi_{mag}=0.8$ after the threshold is reached.
All experiments are performed on the platform with Intel(R) Xeon Gold 5218R CPU and NVIDIA GeForce RTX 3090 GPU.

\subsection{Fidelity and Capacity}
\subsubsection{Analysis of Fidelity}
Fidelity measures how much the speech quality is affected when watermarking is integrated into the diffusion-based generation pipeline. 
Table~\ref{tab_fidelity} reports fidelity results under a fixed payload of 100 bps (watermark length $l=100$ for each 1-second clip) using four speech-fidelity metrics (STOI, PESQ, SSIM, and MCD) across three diffusion vocoders (DiffWave, PriorGrad, WaveGrad) and three datasets (LJSpeech, LibriTTS, LibriSpeech). To disentangle where fidelity degradation comes from, we conduct three pairwise comparisons: i) \textit{Generated$\leftrightarrow$Natural}, comparing unwatermarked generated speech with the corresponding clean speech to measure the generation-induced fidelity gap; ii) \textit{Watermarked $\leftrightarrow$Generated}, comparing watermarked and unwatermarked outputs from the same vocoder to quantify the incremental impact of watermarking; iii) \textit{Watermarked$\leftrightarrow$Natural}, comparing the final watermarked output with clean speech to reflect the overall gap. 

A key observation is that diffusion synthesis itself can introduce a substantial fidelity gap, especially on multi-speaker data. For example, with DiffWave on LibriSpeech, the MCD values of Generated$\leftrightarrow$Natural (15.2482) and Watermarked$\leftrightarrow$Natural (15.3361) are highly consistent, suggesting that most of the overall fidelity gap is attributable to the diffusion generation process. Meanwhile, the corresponding Watermarked$\leftrightarrow$Generated MCD is only 1.1176, indicating that enabling watermarking causes only a limited additional change relative to the unwatermarked generated baseline. A similar pattern is observed on WaveGrad with LibriSpeech: the MCD for Generated$\leftrightarrow$Natural is 16.2426, and the MCD for Watermarked$\leftrightarrow$Natural is 16.8436, whereas the Watermarked$\leftrightarrow$ Generated comparison yields a much smaller MCD of 3.0281. 
Taken together, these results imply that the observed fidelity degradation is largely attributable to the vocoder synthesis, while TriniMark introduces a comparatively small incremental impact. 

Accordingly, in the remainder of our experiments, we primarily report Watermarked$\leftrightarrow$Generated as the fidelity indicator, since it isolates the incremental distortion introduced by watermarking from the intrinsic synthesis gap of the diffusion vocoder.

\begin{table*}[t]
\centering
\renewcommand\arraystretch{0.5}
\setlength{\belowcaptionskip}{-0.1cm}
\caption{Fidelity of TriniMark With Various Datasets in Different DMs. ↑/↓ indicates a higher/lower value is more desirable.}
\vspace{-0.3cm}
\resizebox{0.9\linewidth}{!}{
\begin{tabular}{cccccccccc}
\toprule
\multirow{2}{*}{DiffWave} & \multicolumn{3}{c}{\textit{Generated $\leftrightarrow$ Natural}} & \multicolumn{3}{c}{\textit{Watermarked $\leftrightarrow$ Generated}} & \multicolumn{3}{c}{\textit{Watermarked $\leftrightarrow$ Natural}} \\
\cmidrule{2-10}
& LJSpeech    & LibriTTS   & \multicolumn{1}{c|}{LibriSpeech} & LJSpeech  & LibriTTS   & \multicolumn{1}{c|}{LibriSpeech}  & LJSpeech   & LibriTTS   & LibriSpeech   \\
\midrule
STOI↑   & 0.9655 & 0.9337 & \multicolumn{1}{c|}{0.9176} & 0.9621 & 0.9386 & \multicolumn{1}{c|}{0.9290} & 0.9622 & 0.9542 & 0.9416  \\
PESQ↑   & 3.5120 & 2.8156 & \multicolumn{1}{c|}{2.7788} & 3.1335 & 2.7773 & \multicolumn{1}{c|}{2.6225} & 3.1265 & 2.7860 & 2.6265  \\
SSIM↑   & 0.8453 & 0.8025 & \multicolumn{1}{c|}{0.6699} & 0.9174 & 0.8945 & \multicolumn{1}{c|}{0.8392}  & 0.8567 & 0.8324 & 0.7205  \\
MCD↓    & 6.2794 & 6.0407 & \multicolumn{1}{c|}{15.2482} & 1.2644 & 1.2127 & \multicolumn{1}{c|}{1.1176} & 6.0290 & 6.0031 & 15.3361 \\
ACC↑    & N/A    & N/A    & \multicolumn{1}{c|}{N/A}    & 0.9843 & 0.9588 & \multicolumn{1}{c|}{0.9806}  & 0.9840 & 0.9588 & 0.9809 \\
\midrule
\multirow{2}{*}{PriorGrad} & \multicolumn{3}{c}{\textit{Generated $\leftrightarrow$ Natural}} & \multicolumn{3}{c}{\textit{Watermarked $\leftrightarrow$ Generated}} & \multicolumn{3}{c}{\textit{Watermarked $\leftrightarrow$ Natural}} \\
\cmidrule{2-10}
& LJSpeech    & LibriTTS    & \multicolumn{1}{c|}{LibriSpeech} & LJSpeech    & LibriTTS    & \multicolumn{1}{c|}{LibriSpeech}  & LJSpeech    & LibriTTS   & LibriSpeech   \\
\midrule
STOI↑   & 0.9722 & 0.9463  & \multicolumn{1}{c|}{0.8970} & 0.9424 & 0.9186 & \multicolumn{1}{c|}{0.9201} & 0.9585 & 0.9421 & 0.9480  \\
PESQ↑   & 3.8875 & 2.0509  & \multicolumn{1}{c|}{1.9728} & 2.1856 & 1.7864 & \multicolumn{1}{c|}{2.3841} & 2.1928 & 1.7870 & 2.3910  \\
SSIM↑   & 0.9032 & 0.7130  & \multicolumn{1}{c|}{0.5642} & 0.9177 & 0.8876 & \multicolumn{1}{c|}{0.8497}  & 0.8943 & 0.8652 & 0.7885 \\
MCD↓    & 5.5146 & 16.4618 & \multicolumn{1}{c|}{34.8790} & 2.0892 & 2.2353 & \multicolumn{1}{c|}{1.9604} & 5.6031 & 7.2721 & 12.0343 \\
ACC↑    & N/A    & N/A     & \multicolumn{1}{c|}{N/A}    & 0.9811 & 0.9981 & \multicolumn{1}{c|}{0.9986}  & 0.9799 & 0.9981 & 0.9987 \\
\midrule
\multirow{2}{*}{WaveGrad} & \multicolumn{3}{c}{\textit{Generated $\leftrightarrow$ Natural}} & \multicolumn{3}{c}{\textit{Watermarked $\leftrightarrow$ Generated}} & \multicolumn{3}{c}{\textit{Watermarked $\leftrightarrow$ Natural}} \\
\cmidrule{2-10}
& LJSpeech    & LibriTTS    & \multicolumn{1}{c|}{LibriSpeech} & LJSpeech    & LibriTTS    & \multicolumn{1}{c|}{LibriSpeech}  & LJSpeech    & LibriTTS   & LibriSpeech   \\
\midrule
STOI↑   & 0.9363 & 0.8996  & \multicolumn{1}{c|}{0.8792} & 0.8978 & 0.8677 & \multicolumn{1}{c|}{0.8349} & 0.9169 & 0.8816 & 0.8598  \\
PESQ↑   & 2.2339 & 1.7483  & \multicolumn{1}{c|}{1.9555} & 2.0913 & 1.7754 & \multicolumn{1}{c|}{1.8016}  & 2.0926 & 1.7796 & 1.8021 \\
SSIM↑   & 0.7448 & 0.7533  & \multicolumn{1}{c|}{0.6434} & 0.8426 & 0.8058 & \multicolumn{1}{c|}{0.7168} & 0.7786 & 0.7503 & 0.6287  \\
MCD↓    & 8.6023 & 5.9581 & \multicolumn{1}{c|}{16.2426} & 1.9275 & 2.2932 & \multicolumn{1}{c|}{3.0281} & 6.4706 & 4.7066 & 16.8436 \\
ACC↑    & N/A    & N/A     & \multicolumn{1}{c|}{N/A}    & 0.9821 & 0.9702 & \multicolumn{1}{c|}{0.9316}  & 0.9813 & 0.9693 & 0.9300 \\
\bottomrule
\end{tabular}
}
\vspace{-0.4cm}
\label{tab_fidelity}
\end{table*}

\begin{figure*}
    \centering
    \includegraphics[width=0.95\textwidth]{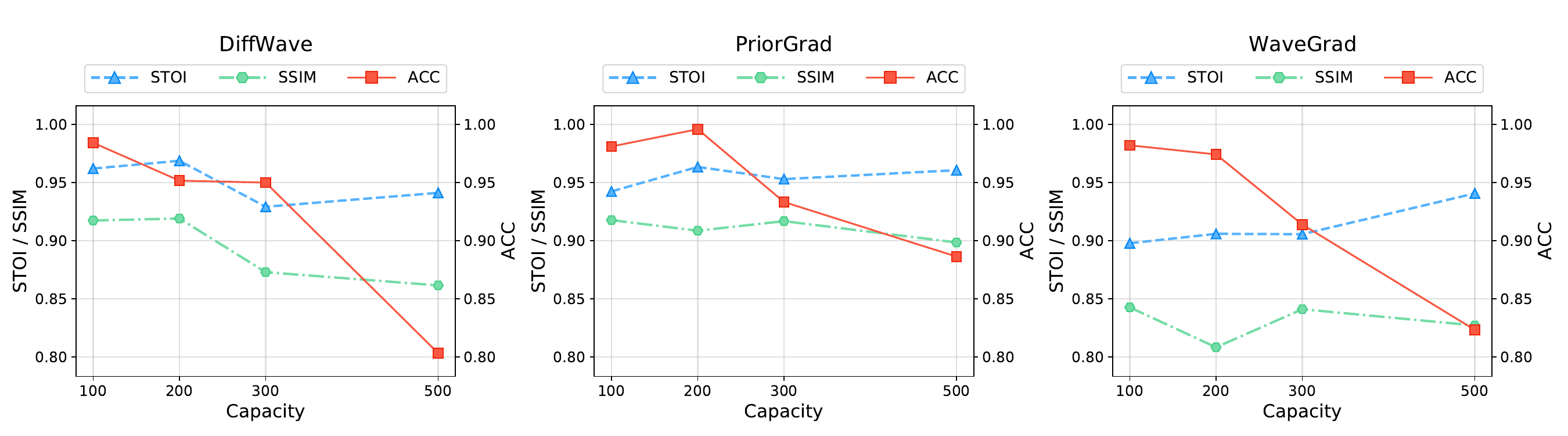}
    \vspace{-0.3cm}
    \caption{Capacity–quality–accuracy Trade-off of TriniMark across Three Diffusion Vocoders on LJSpeech.}
    \label{fig:capacity}
    \vspace{-0.2cm}
\end{figure*}


\begin{table}[]
\centering
\caption{Fidelity-Capacity Comparison With SOTA Methods. ↑/↓ indicates a higher/lower value is more desirable.}
\vspace{-0.3cm}
\resizebox{0.98\linewidth}{!}{
\begin{tabular}{ccccccc}
\toprule
Methods & \makecell{Capacity\\(bps)} & STOI↑  & PESQ↑  & SSIM↑  & MCD↓  & \multicolumn{1}{c}{ACC↑} \\
\midrule
AudioSeal~\cite{roman2024audioseal}       & 16   & 0.9972 & 4.3890 & \textbf{0.9866} & \textbf{0.0090} & \textbf{1.0000} \\
HiFi-GANw~\cite{cheng2024hifi}  & 20  & 0.9414 & 2.5862 & 0.9447 & 4.8472 & 0.9893 \\
WavMark~\cite{chen2023wavmark}         & 32   & \textbf{0.9997} & \textbf{4.4625} & 0.9690 & 2.0504 & \textbf{1.0000} \\
TimbreWM~\cite{liu2024timebre}        & 100  & 0.9853 & 4.0371 & 0.9388 & 3.1715 & 0.9998 \\
\rowcolor[HTML]{E8F7F2} 
TriniMark(DW)  & 100  & 0.9621 & 3.1335 & 0.9174 & 1.2644 & 0.9843 \\
\rowcolor[HTML]{E8F7F2} 
TriniMark(PG) & 100  & 0.9424 & 2.1856 & 0.9177 & 2.0892 & 0.9811 \\
\rowcolor[HTML]{E8F7F2} 
TriniMark(WG) & 100  & 0.8978 & 2.0913 & 0.8426 & 1.9275 & 0.9821 \\
\midrule
TimbreWM~\cite{liu2024timebre}        & 500  & 0.9987 & 4.6322 & 0.9990 & 0.9811 & 0.4999 \\
\rowcolor[HTML]{E8F7F2} 
TriniMark(DW)  & 500  & 0.9412 & 2.1151 & 0.8615 & 1.4438 & \textbf{0.8031} \\
\rowcolor[HTML]{E8F7F2} 
TriniMark(PG) & 500  & 0.9607 & 2.5963 & 0.8985 & 2.0893 & \textbf{0.8863} \\
\rowcolor[HTML]{E8F7F2} 
TriniMark(WG) & 500  & 0.9407 & 1.9365 & 0.8270 & 2.1455 & \textbf{0.8232} \\
\bottomrule
\end{tabular}
}
\vspace{-0.5cm}
\label{tab_cmp_f&c}
\end{table}

 
\begin{table*}[t]
\centering
\caption{Robustness of TriniMark Against Individual Attacks across Datasets and Diffusion Vocoders. 
}
\vspace{-0.3cm}
\resizebox{0.95\textwidth}{!}{
\begin{tabular}{ccccccccccccccc}
 \toprule
 \multirow{2}{*}{DMs}     & \multirow{2}{*}{Dataset}  & \multirow{2}{*}{} & \multicolumn{4}{c}{Gaussian Noise} & PN & LP & BP & \multicolumn{2}{c}{Suppression} & Echo & Stretch & Dither    \\
 \cmidrule{4-15}
 &&& 5 dB & 10 dB & 15 dB & 20 dB & 0.5 & 3k & 0.5-8k & front & behind & default & 2$\times$ & default\\
 \midrule
 & \multirow{3}{*}{LJSpeech} & STOI↑ & 0.8351 & 0.9028  & 0.9491  & 0.9762 & 0.8498 & 0.9997 & 0.8619 & 0.4136 & 0.5442  & 0.7777 & 0.9999 & 1.0000  \\
 & & PESQ↑ & 3.1257 & 3.1320 & 3.1245 & 3.1270 & 3.1282 & 3.1356 & 3.1326 & 3.1227 & 3.1230  & 3.1347 & 3.1341 & 3.1299  \\
 & & ACC↑ & 0.6777 & 0.7761 & 0.8777 & 0.9458 & 0.8487 & 0.9335 & 0.9838 & 0.8649 & 0.9527  & 0.9483 & 0.9809 & 0.9829  \\
 \cmidrule{2-15}
  & \multirow{3}{*}{LibriTTS} & STOI↑ & 0.8372 & 0.8943 & 0.9362 & 0.9639 & 0.8205 & 0.9998 & 0.8843 & 0.4146 & 0.5387 & 0.7728 & 0.9999 & 1.0000  \\
  DiffWave & & PESQ↑ & 2.7822 & 2.7831 & 2.7772 & 2.7837 & 2.7786 & 2.7834 & 2.7757 & 2.7815 & 2.7837 & 2.7792 & 2.7722 & 2.7798 \\
  & & ACC↑ & 0.6868 & 0.7769 & 0.8637 & 0.9207 & 0.7872 & 0.8954 & 0.9582 & 0.8371 & 0.9027 & 0.9056 & 0.9524 & 0.9589  \\
 \cmidrule{2-15}
  & \multirow{3}{*}{LibriSpeech} & STOI↑ & 0.8137 & 0.8635 & 0.9019 & 0.9290 & 0.7864 & 0.9997 & 0.8969 & 0.5336 & 0.4559 & 0.7034 & 0.9999 & 0.9999  \\
  & & PESQ↑ & 2.6197 & 2.6195 & 2.6139 & 2.6229 & 2.6141 & 2.6221 & 2.6205 & 2.6171 & 2.6191 & 2.6182 & 2.6203 & 2.6206 \\
  & & ACC↑ & 0.6880 & 0.7769 & 0.8697 & 0.9324 & 0.8201 & 0.8812 & 0.9807 & 0.9655 & 0.8223 & 0.8646 & 0.9747 & 0.9810  \\
 \midrule
 & \multirow{3}{*}{LJSpeech} & STOI↑ & 0.8638 & 0.9249 & 0.9625 & 0.9832 & 0.9061 & 0.9997 & 0.8615 & 0.4219 & 0.5298  & 0.7525 & 0.9999 & 1.0000  \\
 & & PESQ↑ & 2.1875 & 2.1861 & 2.1815 & 2.1853 & 2.1884 & 2.1853 & 2.1907 & 2.1912  & 2.1880  & 2.1875 & 2.1902 & 2.1874  \\
 &  & ACC↑ & 0.9142 & 0.9585 & 0.9738 & 0.9788 & 0.8797 & 0.9769 & 0.9694 & 0.9127  & 0.9503   & 0.9347 & 0.9807 & 0.9797 \\
 \cmidrule{2-15}
 & \multirow{3}{*}{LibriTTS} & STOI↑ & 0.8588 & 0.9159 & 0.9540 & 0.9766 & 0.9108 & 0.9987 & 0.8800 & 0.4252 & 0.5247 & 0.7646 & 0.9991 & 0.9998  \\
 PriorGrad & & PESQ↑ & 2.1766 & 2.1756 & 2.1731 & 2.1758 & 1.7879 & 2.1741 & 2.1747 & 2.1773 & 2.1735 & 2.1750 & 2.1757 & 1.7872 \\
 & & ACC↑ & 0.8738 & 0.9388 & 0.9663 & 0.9753 & 0.9792 & 0.9756 & 0.9783 & 0.8859 & 0.9531 & 0.9487 & 0.9788 & 0.9982 \\
 \cmidrule{2-15}
 & \multirow{3}{*}{LibriSpeech} & STOI↑ & 0.8236 & 0.8801 & 0.9224 & 0.9525 & 0.8745 & 0.9998 & 0.8931 & 0.6946 & 0.2175 & 0.5843 & 0.9999 & 1.0000  \\
 & & PESQ↑ & 2.3829 & 2.3905 & 2.3918 & 2.3903 & 2.3880 & 2.3898 & 2.3891 & 2.3915 & 2.3887 & 2.3916 & 2.3855 & 2.3901 \\
 & & ACC↑ & 0.7498 & 0.8609 & 0.9453 & 0.9846 & 0.9175 & 0.9966 & 0.9988 & 0.9540 & 0.9972 & 0.9703 & 0.9985 & 0.9987 \\
 \midrule
 & \multirow{3}{*}{LJSpeech}    & STOI↑ & 0.7911 & 0.8722 & 0.9300 & 0.9647 & 0.7995 & 0.8553 & 0.7993 & 0.4059 & 0.5326 & 0.7449 & 0.9967 & 0.9968 \\
 & & PESQ↑ & 2.0943 & 2.0966 & 2.1026 & 2.0992 & 2.1002 & 2.1003 & 2.0893 & 2.0967 & 2.0970 & 2.0976 & 2.0988 & 2.0969 \\
 & & ACC↑  & 0.7562 & 0.8593 & 0.9316 & 0.9644 & 0.9163 & 0.8995 & 0.9818 & 0.8718 & 0.9623 & 0.9594 & 0.9805 & 0.9818 \\                      
 \cmidrule{2-15}
 & \multirow{3}{*}{LibriTTS}    & STOI↑ & 0.8174 & 0.8842 & 0.9327 & 0.9629 & 0.7746 & 0.9666 & 0.8328 & 0.4083 & 0.5350 & 0.7684 & 0.9985 & 0.9985 \\
 WaveGrad & & PESQ↑ & 1.7787 & 1.7724 & 1.7728 & 1.7788 & 1.7733 & 1.7750 & 1.7739 & 1.7749 & 1.7732 & 1.7776 & 1.7810 & 1.7774 \\
 & & ACC↑  & 0.7889 & 0.8795 & 0.9350 & 0.9587 & 0.8968 & 0.9073 & 0.9698 & 0.8626 & 0.9409 & 0.9375 & 0.9706 & 0.9705 \\               
 \cmidrule{2-15}
 & \multirow{3}{*}{LibriSpeech} & STOI↑ & 0.7454 & 0.7921 & 0.8330 & 0.8692 & 0.7093 & 0.9847 & 0.8673 & 0.5893 & 0.3262 & 0.5893 & 0.9986 & 0.9986 \\
 & & PESQ↑ & 1.8076 & 1.8064 & 1.8111 & 1.8064 & 1.8043 & 1.7985 & 1.8034 & 1.8008 & 1.8017 & 1.8057 & 1.8028 & 1.8056 \\
 & & ACC↑  & 0.7290 & 0.8161 & 0.8771 & 0.9124 & 0.8179 & 0.8523 & 0.9312 & 0.8655 & 0.8292 & 0.8515 & 0.9302 & 0.9326 \\ 
 \bottomrule
\end{tabular}
}
\vspace{-0.3cm}
\label{tab_robust_indi}
\end{table*}

\subsubsection{Capacity for Scalable Traceability}
Capacity directly determines the scalability of traceability, especially for user-level tracing where each user should be assigned a distinct watermark message. Figure~\ref{fig:capacity} summarizes the capacity–performance trade-off on three diffusion vocoders by varying the payload from 100 to 500 bps, and reporting speech fidelity (STOI/SSIM) and decoding accuracy (ACC), all computed between the watermarked speech and its unwatermarked generated counterpart.
As capacity increases, ACC decreases monotonically, while STOI and SSIM vary more mildly and remain relatively stable until the highest-capacity regime. In particular, ACC stays high at 100-300 bps but drops more noticeably at 500 bps; the degradation is most pronounced on DiffWave (down to 80.31\%) and less severe on PriorGrad (down to 88.63\%), with WaveGrad exhibiting a similar downward trend and becoming more sensitive as capacity approaches 500 bps. Overall, these results suggest that increasing capacity mainly stresses decodability, whereas the impact on perceived speech quality is comparatively limited, supporting TriniMark’s scalability for user-level traceability under practical payload budgets.




\underline{\textit{However, why high capacity matters for traceability?}} Our design supports variable-watermark assignment, enabling different users to receive different messages during inference. In principle, if each user is assigned a unique $l$-bit binary watermark, the maximum number of distinct user identities that can be represented is $2^l$. Therefore, at $l=500$, the raw identifier space is $2^{500}$, which is astronomically large. However, this number should be interpreted as a theoretical upper bound: in practice, the effective user capacity is determined by error rates, decision thresholds, and whether error-correction is applied. High payload also provides two practical benefits beyond sheer ID space: \ding{182} \textit{Error tolerance via error correcting code.} With a 500-bit budget, even when decoding incurs bit errors, we can allocate part of the payload to error correcting redundancy to make user identification robust under distortions. In other words, higher 
$l$ offers a larger “coding margin” to trade bits for reliability without collapsing the available user space.
\ding{183} \textit{Room for collusion-resistant fingerprinting.} If we further consider user collusion attacks, a large payload enables embedding anti-collusion fingerprinting codes \cite{tardos2008optimal} rather than plain random IDs. Such codes require non-trivial code lengths to support many users under a given collusion size; thus, the 500-bit regime provides meaningful headroom to incorporate collusion resistance while still retaining a large user population.

\vspace{-0.2cm}
\subsubsection{Comparison of Fidelity and Capacity With SOTA Methods}
Table~\ref{tab_cmp_f&c} compares TriniMark with representative SOTA speech watermarking baselines in terms of capacity and fidelity. AudioSeal~\cite{roman2024audioseal}, WavMark~\cite{chen2023wavmark}, and TimbreWM~\cite{liu2024timebre} are post-hoc methods that operate on the final waveform and therefore mainly support content-level provenance, while HiFi-GANw~\cite{cheng2024hifi} is a generative baseline with variable watermarking. The key limitation shared by these baselines is that their payloads (typically 16-32 bps) are too small to support scalable multi-user or multi-content traceability. To stress-test capacity, we further evaluate the largest capacity baseline, TimbreWM, at 500 bps. As shown in Table \ref{tab_cmp_f&c}, the ACC of Timbre drops to 0.4999, indicating that it can no longer provide reliable extraction at this payload. In contrast, TriniMark remains decodable at 500 bps, with ACC above 80\% across all three diffusion vocoders. This confirms that TriniMark supports substantially higher payloads.

Regarding fidelity, TriniMark does not always achieve the best PESQ, but it preserves competitive STOI and SSIM. At 100 bps, TriniMark yields notably better MCD than most methods except AudioSeal. AudioSeal reports an extremely low MCD (0.0090), but its payload is only 16 bps. At 500 bps, TriniMark exhibits a predictable fidelity–capacity trade-off. At the subjective level, we further listen to the generated speech to assess perceptual quality. At 100 bps, the watermarked speech is perceptually indistinguishable from its unwatermarked counterpart in casual listening. When increasing the payload to 500 bps, slight background noise can be noticed in some samples. Nevertheless, when the listener is not provided with the original reference and only hears the generated watermarked speech, such artifacts are generally not perceived as abrupt or overly distracting.




\begin{table*}[t]
\centering
\renewcommand\arraystretch{0.1}
\setlength{\tabcolsep}{3mm}
\caption{Robustness of TriniMark Against Compound Attacks across Datasets and Diffusion Vocoders}
\vspace{-0.3cm}
\small
\begin{adjustbox}{max width=\textwidth}
\begin{tabular}{ccccccccc}
\toprule
\multirow{2}{*}{Datasets} & & \multicolumn{7}{c}{DiffWave} \\
\cmidrule{3-9}
                             &      & GN+BP  & GN+Echo & GN+Dither & GN+PN & PN+BP  & PN+Echo & PN+Dither \\
\midrule
\multirow{3}{*}{LJSpeech}    & STOI↑ & 0.8453 & 0.7555 & 0.9762 & 0.8445  & 0.7853 & 0.6764 & 0.8495  \\
                             & PESQ↑ & 3.1356 & 3.1254 & 3.1269 & 3.1340  & 3.1264 & 3.1323 & 3.1261  \\
                             & ACC↑  & 0.9449 & 0.9071 & 0.9442 & 0.8285  & 0.8524 & 0.8117 & 0.8482  \\
\midrule
\multirow{3}{*}{LibriTTS}    & STOI↑ & 0.8522 & 0.7477 & 0.9640 & 0.8162  & 0.7612 & 0.6533 & 0.8204  \\
                             & PESQ↑ & 2.7796 & 2.7777 & 2.7846 & 2.7779  & 2.7800 & 2.7804 & 2.7779  \\
                             & ACC↑  & 0.9197 & 0.8694 & 0.9207 & 0.7757  & 0.7957 & 0.7542 & 0.7885  \\
\midrule
\multirow{3}{*}{LibriSpeech} & STOI↑ & 0.8354 & 0.6858 & 0.9292 & 0.7826  & 0.7375 & 0.6009 & 0.7867  \\
                             & PESQ↑ & 2.6161 & 2.6210 & 2.6153 & 2.6218  & 2.6168 & 2.6204 & 2.6173  \\
                             & ACC↑  & 0.9324 & 0.8132 & 0.9327 & 0.8023  & 0.8226 & 0.7168 & 0.8197  \\
\midrule
\multirow{2}{*}{Datasets} & & \multicolumn{7}{c}{PriorGrad} \\
\cmidrule{3-9}
                             & & GN+BP  & GN+Echo & GN+Dither & GN+PN & PN+BP  & PN+Echo & PN+Dither \\
\midrule
\multirow{3}{*}{LJSpeech}    & STOI↑ & 0.8497 & 0.7452 & 0.9830 & 0.9012  & 0.8194 & 0.6968 & 0.9052 \\
                             & PESQ↑ & 2.1873 & 2.1925 & 2.1856 & 2.1896  & 2.1862 & 2.1907 & 2.1892 \\
                             & ACC↑  & 0.9670 & 0.9334 & 0.9787 & 0.8796  & 0.9355 & 0.8313 & 0.8820 \\
\midrule
\multirow{3}{*}{LibriTTS}    & STOI↑ & 0.8616 & 0.7387 & 0.9804 & 0.9030  & 0.8246 & 0.8244 & 0.9099 \\
                             & PESQ↑ & 1.7876 & 1.7858 & 1.7881 & 1.7866  & 1.7866 & 1.7889 & 1.7870 \\
                             & ACC↑  & 0.9972 & 0.9856 & 0.9973 & 0.9772  & 0.9806 & 0.9802 & 0.9788 \\
\midrule
\multirow{3}{*}{LibriSpeech} & STOI↑ & 0.8515 & 0.5618 & 0.9525 & 0.8671  & 0.8039 & 0.5372 & 0.8737 \\
                             & PESQ↑ & 2.3959 & 2.3924 & 2.3923 & 2.3881  & 2.3912 & 2.3920 & 2.3907 \\
                             & ACC↑  & 0.9850 & 0.9342 & 0.9846 & 0.9011  & 0.9192 & 0.8423 & 0.9171 \\
\midrule
\multirow{2}{*}{Datasets}    & & \multicolumn{7}{c}{WaveGrad} \\
\cmidrule{3-9}
                             &  & GN+BP  & GN+Echo & GN+Dither & GN+PN & PN+BP  & PN+Echo & PN+Dither \\
\midrule
\multirow{3}{*}{LJSpeech}    & STOI↑ & 0.7811 & 0.7226 & 0.9646 & 0.7933  & 0.7878 & 0.7207 & 0.9478  \\
                             & PESQ↑ & 2.0871 & 2.0982 & 2.0992 & 2.0995  & 2.1081 & 2.1066 & 2.0970  \\
                             & ACC↑  & 0.9665 & 0.9380 & 0.9643 & 0.9035  & 0.9785 & 0.9536 & 0.9786  \\
\midrule
\multirow{3}{*}{LibriTTS}    & STOI↑ & 0.8074 & 0.7453 & 0.9631 & 0.7713  & 0.6165 & 0.7728 & 0.6145  \\
                             & PESQ↑ & 1.7768 & 1.7735 & 1.7753 & 1.7776  & 1.7744 & 1.7763 & 1.7747  \\
                             & ACC↑  & 0.9586 & 0.9259 & 0.9586 & 0.8886  & 0.8615 & 0.8968 & 0.8599  \\
\midrule
\multirow{3}{*}{LibriSpeech} & STOI↑ & 0.7591 & 0.5579 & 0.8685 & 0.7078  & 0.6527 & 0.4833 & 0.7103  \\
                             & PESQ↑ & 1.8066 & 1.8034 & 1.8038 & 1.8086  & 1.7987 & 1.8081 & 1.8018  \\
                             & ACC↑  & 0.9119 & 0.8271 & 0.9104 & 0.8085  & 0.8212 & 0.7430 & 0.8168  \\
                             \bottomrule
\end{tabular}
\end{adjustbox}
\vspace{-0.3cm}
\label{tab_robust_compo}
\end{table*}


\subsection{Robustness}
\subsubsection{Analysis of Robustness Against Individual Attacks}
\label{sec.indiviual_attack}
To evaluate robustness under realistic distribution and post-processing, we consider a suite of individual attacks that can be applied to released waveforms. These attacks fall into four categories: (i) additive noise: Gaussian noise (GN) with SNR of 5, 10, 15, and 20 dB, and pink noise (PN) with a factor of 0.5; (ii) filtering operations: low-pass filtering (LP) with a threshold of 3 kHz, and band-pass filtering (BP) with a 0.5-8 kHz threshold; (iii) channel effects: suppression of front and behind, and echo with default setting in \cite{roman2024audioseal}; (iv) desynchronization: time stretching with an intensity of 2, and dither with default setting in Torchaudio. Table \ref{tab_robust_indi} reports the robustness of TriniMark under the individual attacks described above, measured by speech fidelity (STOI/PESQ) and watermark extracting accuracy (ACC), across three datasets and three diffusion vocoders.

TriniMark exhibits strong robustness under a wide range of individual attacks, particularly under band-pass filtering, time stretching, and dithering. In these cases, the watermark extracting accuracy exceeds 95\% in almost all settings. The most challenging cases are strong Gaussian noise and spectral suppression, because they directly corrupt or remove watermark-carrying cues and effectively reduce the usable SNR for reliable bit recovery. Nevertheless, TriniMark remains robust even under these harsher conditions. For instance, under 5 dB Gaussian noise, PriorGrad on LJSpeech still achieves an ACC of 91.42\%; under spectral suppression on LJSpeech, WaveGrad attains ACC values of 87.18\% (front) and 96.23\% (behind). 

Among the three diffusion vocoders presented in Table \ref{tab_robust_indi}, PriorGrad is consistently the most robust under strong distortions, particularly under Gaussian noise. We attribute this advantage to its use of data-driven statistical priors during denoising, which provides a stronger inductive bias for recovering clean speech features and thus more effectively suppresses noise perturbations along the diffusion trajectory.

\begin{table*}[t]
\centering
\large
\caption{Comparison of Robustness Against Individual Attacks with SOTA Methods. ↑ indicates a higher value is more desirable. \\ The best results are marked in \textbf{bold} and the second best results are \underline{underline}.}
\vspace{-0.3cm}
\resizebox{0.98\textwidth}{!}{
\begin{tabular}{ccccccccccccccccccc}
\toprule
\multirow{2}{*}{} & \multicolumn{6}{c}{Gaussian Noise} & \multicolumn{3}{c}{Pink Noise} & \multicolumn{6}{c}{Suppression} & \multicolumn{3}{c}{Echo} \\
\cmidrule{2-19}
& \multicolumn{3}{c}{10 dB} & \multicolumn{3}{c|}{20 dB} & \multicolumn{3}{c|}{0.5}& \multicolumn{3}{c}{Front} & \multicolumn{3}{c|}{Behind} & \multicolumn{3}{c}{Default} \\
\midrule
& STOI↑  & PESQ↑  & \multicolumn{1}{c|}{ACC↑}   & STOI↑  & PESQ↑  & \multicolumn{1}{c|}{ACC↑}   & STOI↑  & PESQ↑  & \multicolumn{1}{c|}{ACC↑}   & STOI↑  & PESQ↑  & \multicolumn{1}{c|}{ACC↑}   & STOI↑  & PESQ↑  & \multicolumn{1}{c|}{ACC↑}  & STOI↑  & \multicolumn{1}{c}{PESQ↑} & \multicolumn{1}{c}{ACC↑} \\
\midrule
WavMark~\cite{chen2023wavmark} & 0.8898 & 1.2217 & \multicolumn{1}{c|}{0.5214} & 0.9733 & 2.1236 & \multicolumn{1}{c|}{0.6523} & 0.8673 & 1.3019 & \multicolumn{1}{c|}{0.6924}& 0.4185 & 1.7090 & \multicolumn{1}{c|}{\underline{0.9797}} & 0.5135 & 1.7466 & \multicolumn{1}{c|}{0.9713} & 0.6122 & 1.3716 & 0.8668 \\
AudioSeal~\cite{roman2024audioseal} & 0.9110 & 1.0995 & \multicolumn{1}{c|}{0.6086} & 0.9789 & 1.5987 & \multicolumn{1}{c|}{0.6600} & 0.9185 & 1.3537 & \multicolumn{1}{c|}{0.6571} & 0.4150 & 1.0916 & \multicolumn{1}{c|}{0.7226} & 0.5348 & 1.1661 & \multicolumn{1}{c|}{0.8925} & 0.7563 & 1.1845 & 0.7277 \\
TimbreWM~\cite{liu2024timebre} & 0.9136 & 1.3347 & \multicolumn{1}{c|}{0.6335} & 0.9812 & 2.7424 & \multicolumn{1}{c|}{0.8154} & 0.8473 & 1.3773 & \multicolumn{1}{c|}{0.7282} & 0.4135 & 1.8149 & \multicolumn{1}{c|}{\textbf{0.9888}} & 0.5331 & 1.8155 & \multicolumn{1}{c|}{\textbf{0.9814}} & 0.7559 & 1.4720 & 0.5818 \\
HiFi-GANw~\cite{cheng2024hifi} & 0.9254 & 1.1030 & \multicolumn{1}{c|}{0.7957} & 0.9853 & 1.6113 & \multicolumn{1}{c|}{0.8208} & 0.8473 & 1.0793 & \multicolumn{1}{c|}{\textbf{0.9588}} & 0.4148 & 1.0941 & \multicolumn{1}{c|}{0.9508} & 0.5377 & 1.1706 & \multicolumn{1}{c|}{\underline{0.9717}} & 0.9873 & 4.5597 & \textbf{0.9829} \\
\rowcolor[HTML]{E8F7F2} 
TriniMark(DW)  & 0.9028 & 3.1320 & \multicolumn{1}{c|}{0.7761} & 0.9762 & 3.1270 & \multicolumn{1}{c|}{0.9458} & 0.8498 & 3.1282 & \multicolumn{1}{c|}{0.8487} & 0.4136 & 3.1227 & \multicolumn{1}{c|}{0.8649} & 0.5442 & 3.1230 & \multicolumn{1}{c|}{0.9527} & 0.7777 & 3.1347 & 0.9483 \\
\rowcolor[HTML]{E8F7F2} 
TriniMark(PG) & 0.9249 & 2.1861 & \multicolumn{1}{c|}{\textbf{0.9585}} & 0.9832 & 2.1853 & \multicolumn{1}{c|}{\textbf{0.9788}} & 0.9061 & 2.1884 & \multicolumn{1}{c|}{0.8797} & 0.4219 & 2.1912 & \multicolumn{1}{c|}{0.9127} & 0.5298 & 2.1880 & \multicolumn{1}{c|}{0.9503} & 0.7525 & 2.1875 & 0.9347 \\
\rowcolor[HTML]{E8F7F2} 
TriniMark(WG) & 0.8722 & 2.0966 & \multicolumn{1}{c|}{\underline{0.8593}} & 0.9647 & 2.0992 & \multicolumn{1}{c|}{\underline{0.9644}} & 0.7995 & 2.1002 & \multicolumn{1}{c|}{\underline{0.9163}} & 0.4059 & 2.0967 & \multicolumn{1}{c|}{0.8718} & 0.5326 & 2.0970 & \multicolumn{1}{c|}{0.9623} & 0.7449 & 2.0976 & \underline{0.9594} \\
\bottomrule
\end{tabular}
}
\vspace{-0.3cm}
\label{tab_robust_cmp_indi}
\end{table*}

\begin{table*}
\setlength{\belowcaptionskip}{-0.01mm}
\centering
\caption{Comparison of Robustness against Compound Attacks with SOTA Methods.↑ indicates a higher value is more desirable. \\ The best results are marked in \textbf{bold} and the second best results are \underline{underline}.}
\vspace{-0.3cm}
\resizebox{0.98\textwidth}{!}{
\begin{tabular}{ccccccccccccccccccc}
\toprule
 & \multicolumn{3}{c}{GN+BP} & \multicolumn{3}{c}{GN+Echo}  & \multicolumn{3}{c}{GN+PN} & \multicolumn{3}{c}{PN+BP} & \multicolumn{3}{c}{PN+Echo}  \\
 \cmidrule{2-16}
  & STOI↑  & PESQ↑  & \multicolumn{1}{c|}{ACC↑} & STOI↑  & PESQ↑  & \multicolumn{1}{c|}{ACC↑} & STOI↑  & PESQ↑  & \multicolumn{1}{c|}{ACC↑} & STOI↑  & PESQ↑  & \multicolumn{1}{c|}{ACC↑} & STOI↑  & PESQ↑  & ACC↑ \\
  \midrule
WavMark~\cite{chen2023wavmark}         & 0.8886 & 1.1139 & \multicolumn{1}{c|}{0.6494}          & 0.5922 & 1.2798 & \multicolumn{1}{c|}{0.5547}          & 0.8601 & 1.2292 & \multicolumn{1}{c|}{0.5957}          & 0.8102 & 1.0785 & \multicolumn{1}{c|}{0.6608}          & 0.5221 & 1.1868 & 0.5617          \\
AudioSeal~\cite{roman2024audioseal}       & 0.8447 & 1.5308 & \multicolumn{1}{c|}{0.6409}          & 0.7400 & 1.1177 & \multicolumn{1}{c|}{0.6305}          & 0.9118 & 1.2292 & \multicolumn{1}{c|}{0.6394}          & 0.8292 & 1.7364 & \multicolumn{1}{c|}{0.6480}          & 0.7099 & 1.0912 & 0.6280          \\
TimbreWM~\cite{liu2024timebre}        & 0.7915 & 1.0868 & \multicolumn{1}{c|}{0.5961}          & 0.6905 & 1.2031 & \multicolumn{1}{c|}{0.5935}          & 0.8090 & 1.1672 & \multicolumn{1}{c|}{0.5907}          & 0.7813 & 1.1786 & \multicolumn{1}{c|}{0.7049}          & 0.6470 & 1.2072 & 0.7675          \\
HiFi-GANw~\cite{cheng2024hifi}        & 0.8004 & 1.1068 & \multicolumn{1}{c|}{0.7894}          & 0.7040 & 1.0539 & \multicolumn{1}{c|}{0.8115}          & 0.7949 & 1.0387 & \multicolumn{1}{c|}{0.7879}          & 0.7607 & 1.2386 & \multicolumn{1}{c|}{0.9340}          & 0.6359 & 1.0438 & \textbf{0.9602}          \\
\rowcolor[HTML]{E8F7F2} 
TriniMark(DW)  & 0.8453 & 3.1356 & \multicolumn{1}{c|}{0.9449}    & 0.7555 & 3.1254 & \multicolumn{1}{c|}{0.9071}    & 0.8445 & 3.1340 & \multicolumn{1}{c|}{0.8285}    & 0.7853 & 3.1264 & \multicolumn{1}{c|}{0.8524}    & 0.6764 & 3.1323 & 0.8117    \\
\rowcolor[HTML]{E8F7F2} 
TriniMark(PG) & 0.8497 & 2.1873 & \multicolumn{1}{c|}{\textbf{0.9670}} & 0.7452 & 2.1925 & \multicolumn{1}{c|}{\underline{0.9334}} & 0.9012 & 2.1896 & \multicolumn{1}{c|}{\underline{0.8796}} & 0.8194 & 2.1862 & \multicolumn{1}{c|}{\underline{0.9355}} & 0.6968 & 2.1907 & 0.8313 \\
\rowcolor[HTML]{E8F7F2} 
TriniMark(WG) & 0.7811 & 2.0871 & \multicolumn{1}{c|}{\underline{0.9665}} & 0.7226 & 2.0982 & \multicolumn{1}{c|}{\textbf{0.9380}} & 0.7933 & 2.0995 & \multicolumn{1}{c|}{\textbf{0.9035}} & 0.7878 & 2.1081 & \multicolumn{1}{c|}{\textbf{0.9785}} & 0.7207 & 2.1066 & \underline{0.9536} \\
\bottomrule
\end{tabular}
}
\vspace{-0.3cm}
\label{tab_robust_cmp_compo}
\end{table*}

\begin{figure*}
     \centering
     \resizebox{0.95\textwidth}{!}{\includegraphics{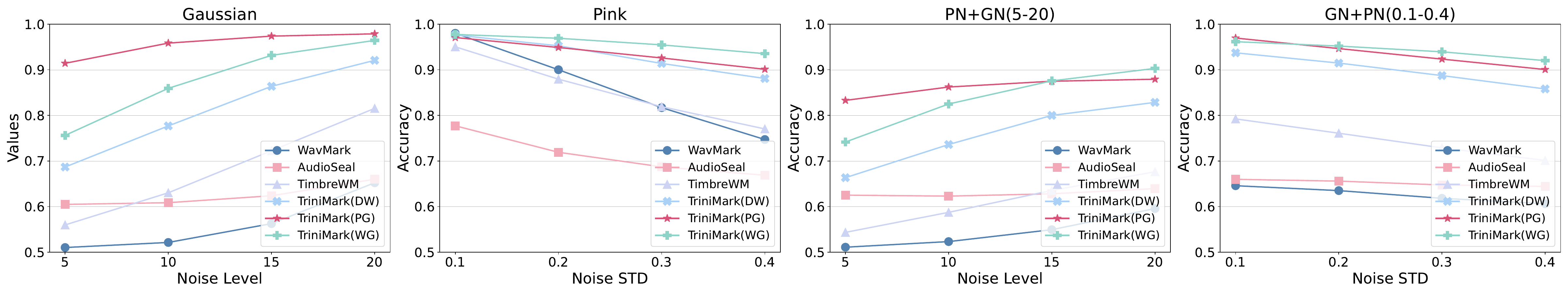}}
     \vspace{-0.3cm}
     \caption{Comparison of Robustness under Noise-level Attacks. For \textit{Gaussian} and \textit{PN+GN}, Gaussian noise levels are 5, 10, 15, and 20 dB. As the noise level decreases, the attack strength increases. For \textit{Pink} and \textit{GN+PN}, noise standard deviations (STD) of pink noise are 0.1, 0.2, 0.3, and 0.4. As the STD increases, the attack strength increases.}
     \label{fig_cmp_robust}
     \vspace{-0.3cm}
\end{figure*}

\vspace{-0.4cm}
\subsubsection{Analysis of Robustness Against Compound Attacks}
\label{sec_ro_comp}

In real-world distribution, a released waveform may undergo more than one post-processing operation. Building on the individual attacks in Section \ref{sec.indiviual_attack}, we therefore evaluate TriniMark under seven compound attacks formed by combining two distortions from the same attack suite, including Gaussian-noise-based and pink-noise-based combinations. 
Table~\ref{tab_robust_compo} shows stable robustness of TriniMark across datasets and diffusion vocoders. The method performs particularly well when noise is paired with band-pass filtering or dithering. For example, under Diffwave, the ACC exceeds 93\% for GN+BP and GN+Dither on both LJSpeech and LibriSpeech; under WaveGrad on LJSpeech, GN+BP and PN+Dither attain ACC values above 96\%. This indicates that TriniMark is highly resilient to such "noise+filtering/dithering" combinations.

The most challenging compound cases typically involve echo or mixed-noise perturbations (GN+PN). Echo effectively mixes the waveform with delayed replicas, thereby introducing temporal overlap and comb-like spectral distortion, while pink noise concentrates interference in low frequencies, so their combination more strongly distorts the speech structure and reduces the effective SNR for reliable watermark extraction. This is most evident on harder multi-speaker data, for example, on LibriSpeech, DiffWave drops to 0.8132 (GN+Echo) and 0.7168 (PN+Echo), while WaveGrad reaches its lowest ACC at 0.7430 (PN+Echo). Despite these severe conditions, TriniMark still maintains usable extraction accuracy rather than collapsing.
Among the three vocoders, PriorGrad is the most robust under compound distortions. Its ACC remains high across most compound settings (e.g., 0.9850 for GN+BP and 0.9846 for GN+Dither on LibriSpeech), and it is notably more stable than DiffWave and WaveGrad on the most challenging echo- and pink-noise-related cases.

\vspace{-0.3cm}
\subsubsection{Comparison of Robustness With SOTA Methods Against Individual Attacks}
We further compare TriniMark with representative SOTA baselines under individual attacks. Table~\ref{tab_robust_cmp_indi} reports robustness results under Gaussian noise, pink noise, spectral suppression, and echo. TriniMark achieves the best robustness under Gaussian noise across the compared methods. In particular, TriniMark on PriorGrad attains ACC of 0.9585 at 10 dB and 0.9788 at 20 dB, surpassing all baselines at the same settings. TriniMark also ranks second under both pink noise and echo: while HiFi-GANw yields the highest ACC, whereas TriniMark follows closely with 91.63\% (pink noise) and 95.94\% (echo). For spectral suppression, although TimbreWM remains the strongest, TriniMark stays competitive; notably, under front suppression, TriniMark achieves 0.8718, about 15\% absolute higher than AudioSeal. These gains stem from TriniMark’s time-domain watermarking design and training strategy: watermark priors are embedded into time-domain features and decoded via temporal-gated convolutions, which supports robust evidence aggregation under waveform perturbations, while variable-watermark training in both stages together with WGFT mitigates pattern memorization and aligns diffusion generation with decodability, improving robustness under distortions.

\subsubsection{Comparison of Robustness With SOTA Methods Against Compound Attacks}
We also compare TriniMark with SOTA baselines under compound attacks. We intentionally include harder pairs involving echo and pink noise, which were already identified as challenging in Section \ref{sec_ro_comp}. As summarized in Table~\ref{tab_robust_cmp_compo}, TriniMark consistently remains among the top performers across the evaluated combinations. It achieves the best (or near-best) robustness on several representative pairs. For more challenging cases involving echo or mixed-noise, TriniMark still maintains strong robustness: under GN+Echo, it achieves an ACC of 93.80\%, and under GN+PN, it achieves 90.35\%. In contrast, most baselines under these settings remain around 60\%–70\% ACC; given their much smaller payload capacities, such degraded decoding reliability makes it difficult to support dependable watermark extraction and, consequently, to distinguish different contents or scale to user-level tracing in practice.
Fig. \ref{fig_cmp_robust} further reveals a clear monotonic trend: robustness improves as Gaussian noise becomes weaker (higher SNR) and degrades as pink-noise STD increases. Across all noise levels, TriniMark variants consistently stay above most baselines, indicating stronger resilience under progressively harsher noise conditions.

\vspace{-0.3cm}
\section{Conclusion}
In this paper, we propose a generative speech watermarking based on fine-grained feature transfer, which establishes a trinity traceability mechanism that simultaneously authenticates three essential dimensions: the generative model, the synthesized speech, and the end-user.
The proposed TriniMark consists of two stages of training.
In the first stage, to achieve efficient transfer of watermark generation to the generative model, a watermark encoder-decoder is designed. 
Specifically, to achieve high-precision watermark extraction, we design a temporal-aware gated convolutional network as the backbone of the watermark decoder.
In the second stage, we further propose a waveform-guided fine-tuning strategy. 
This strategy embeds watermarks into the training data employing the pretrained encoder and jointly optimizes gradients with the pretrained decoder.
At the same time, this fine-tuning strategy enables TriniMark to adapt to arbitrary watermarks with only a single round of training.
Fidelity and capacity experiments demonstrate that TriniMark can generate high-quality watermarked speech even under a high capacity of 500 bps. 
Robustness experiments further verify the superior performance of our method compared to existing approaches when facing both individual and compound attacks.

\bibliographystyle{ACM-Reference-Format}
\bibliography{samples/refer}

\end{document}